\documentstyle[aps,multicol]{revtex}
\draft
\newcommand{\be}{\begin{equation}}
\newcommand{\ee}{\end{equation}}
\newcommand{\bea}{\begin{eqnarray}}
\newcommand{\eea}{\end{eqnarray}}

\def\cF{{\cal F}}
\def\cA{{\cal A}}
\def\Im{{\rm Im}\,}
\def\Re{{\rm Re}\,}
\def\Tr{{\rm Tr}\,}

\begin{document}
  \begin{flushright} \begin{small}
     DTP-MSU/99-01 \\ WU-AP/77/99 \\
     hep-th/9901130
  \end{small} \end{flushright}
\vspace{.5cm}
\begin{center}
{\large \bf $SL(4,R)$ Generating Symmetry in Five--Dimensional Gravity \\
            Coupled to Dilaton and Three--Form}

\vskip.5cm
Chiang-Mei Chen$^{a}$\footnote{Email: chen@grg1.phys.msu.su},
Dmitri V. Gal'tsov$^{a}$\footnote{Email: galtsov@grg.phys.msu.su},
Kei-ichi Maeda$^{b}$\footnote{Email: maeda@mse.waseda.ac.jp}
and Sergei A. Sharakin$^{a}$\footnote{Email: sharakin@grg1.phys.msu.su}

\vskip.2cm
\smallskip
$^{a}${\em Department of Theoretical Physics,
           Moscow State University, 119899, Moscow, Russia}

\smallskip
$^{b}${\em Department of Physics,
           Waseda University, Tokyo 169-8555, Japan}

\end{center}
\vskip.1cm
\date{January 26, 1999}

\begin{abstract}
We give an explicit formulation of the three--dimensional
$SL(4,R)/SO(2,2)\,\sigma$--model representing the five--dimensional
Einstein gravity coupled to the dilaton and the three--form field
for spacetimes with two commuting Killing vector fields. New matrix
representation is obtained which is similar to one found earlier in the
four--dimensional Einstein--Maxwell--Dilaton--Axion (EMDA) theory.
The $SL(4,R)$ symmetry joins a variety of $5D$ solutions of different
physical types including strings, $0$--branes, KK monopoles etc.
interpreting them as duals to the four--dimensional Kerr metric
translated along the fifth coordinate. The symmetry transformations
are used to construct new rotating strings and composite
$(0-1)$--branes endowed with a NUT parameter.
\end{abstract}

\begin{multicols}{2}
\narrowtext

\section{Introduction}
Recent interest to soliton solutions in multidimensional
supergravities stimulates a
search of solution generating techniques relevant to various gravity
coupled systems containing antisymmetric forms. Dimensional reduction
of these theories reveals global symmetries which can be effectively
used for such purposes. But although general structure
of global symmetries in dimensionally reduced supergravities
is well--known~\cite{Ju81,St97},
their practical application requires additional efforts to be done.
One has to construct suitable matrix representations in terms
of which these symmetries, acting non--linearly (and perhaps partly
non--locally) on the set of initial field variables, could be expressed
in a tractable way. In dimension three or in higher dimensions with
suitably truncated metric ans\"atz one is able to express all variables
in terms of scalar fields forming some non--linear $\sigma$--model,
in which case the symmetry transformations are particularly simple.
It has been shown recently~\cite{GaRy98} that all block--diagonal
$p$--brane solutions including black, multiple--center, intersecting
ones as well as fluxbranes and their non--linear superpositions with
charged branes can be produced within a simple $\sigma$--model, the
essential part of which has the same structure as the static
four--dimensional Einstein--Maxwell (EM) system~\cite{NeKr69}.
This model, however, can not be generalized to include rotating or dyon
solutions, contrary to the situation in the stationary four--dimensional
EM or Einstein--Maxwell--Dilaton--Axion (EMDA) systems.
Certain rotating $p$--brane solutions were presented by Cveti\v{c} and
Youm~\cite{CvYo97} apparently hinted by
the Myers and Perry~\cite{MyPe86} multidimensional Kerr--like
metrics, but, to our knowledge, no constructive
derivation of such solutions has been given so far.

As a step further in developing generating techniques for larger
classes of $p$--branes, we suggest here a new matrix representation
for the Maharana--Schwarz symmetry~\cite{MaSc93} of the
five--dimensional gravity coupled to three-form
and dilaton with two commuting Killing vector fields.
The novel feature consists in exploiting an exceptional local
isomorphism between the Maharana--Schwarz group $SO(3,3)$ relevant
to the case and the group $SL(4,R)$. This opens a way to construct
instead of the $6\times 6$ matrix realization a simpler representation
of the symmetry by $4\times 4$ real matrices, which, in turn, may
be regarded as a direct generalization of the $Sp(4,R)$ formulation
of the stationary four--dimensional EMDA theory~\cite{GaKe96}.
This framework naturally incorporates rotating strings as well as
dyon--type solutions which may be interpreted as composite states of strings
and $0$-branes.

Our $\sigma$--model constitutes another example of $4\times 4$ matrix
theories of which it has the maximal dimension of the target space.
The three--dimensional reduction of the 4D EMDA theory with one vector field
corresponds to the $Sp(4,R)/SO(1,2)$ coset which is six-dimensional.
The same theory with two vector fields may be formulated as
$SU(2,2)/U(1,1)$ $\sigma$--model with the eight--dimensional target
space~\cite{GaSh97}.
It is likely that the present case closes the list of theories
whose reduction to three dimensions gives a $4\times 4$ representation,
now for nine--dimensional target space. There is some similarity with
the case of EMDA with two vector fields, in fact the present theory
in four dimensions also has two vector fields, but it is endowed with
an additional scalar modulus. Therefore the target space now is
odd--dimensional and is no more K\"ahler.
Still the matrix structure is very similar to that found previously in
the $Sp(4,R)$ theory, which can be obtained from the present one
by identifying two different $2\times 2$ blocks. (Although the target
space is not K\"ahler, suitable Ernst--like complex potentials can
also be found, in terms of which the target space metrics exhibits a
close similarity to that of the EMDA theory.)
The standard representation of $SL(4,R)$ in terms of the $1\times 3$
block decomposition was applied to five--dimensional dilaton--axion gravity
in \cite{KeYu98}. This formulation,
however, does not make contact with neither $Sp(4,R)$ nor
$SU(2,2)$ theories and thus is less useful for solution generation.

\section{From Five to Four Dimensions}
Our starting point is the five--dimensional action
\be
S_5 = \int d^5x \sqrt{-g_5} \left\{ R_5
    - \frac12 (\partial {\hat\phi})^2
    - \frac{e^{-\alpha\hat\phi}}{12} {\hat H}^2 \right\},
\ee
where $\hat\phi$ is the dilaton, ${\hat H}=d{\hat B}$ is an
antisymmetric three--form.
We perform a two--step reduction to the three--dimensional theory
making first the Kaluza--Klein ans\"atz for the five--dimensional metric
suppressing a spacelike coordinate
\be
ds_5^2 = e^{-4\varphi} \left( dy + \cA_\mu dx^\mu \right)^2
       + e^{2\varphi} g_{\mu\nu} dx^\mu dx^\nu,
\ee
with all fields depending only on $x^\mu$. Then
\bea
\sqrt{-g_5} R_5 &=& \sqrt{-g_4} \Big\{ R_4 - 6 g^{\mu\nu}
  \partial_\mu \varphi \partial_\nu \varphi \nonumber \\
  &-& \frac14 e^{-6\varphi} {\cal F}_{\mu\nu} {\cal F}^{\mu\nu}
  -2\nabla_\mu \left( g^{\mu\nu}\partial_\nu \varphi \right) \Big\}
\eea
where $\cA = \cA_\mu d x^\mu,\; \cF = d \cA$.
The five--dimensional two--form ${\hat B}$ can be decomposed
into the four--dimensional two--form $B$ and the four--dimensional
one--form $A$
\be
{\hat B} = B - d y \wedge A.
\ee
Then the four--dimensional $H$ will include the appropriate
Chern--Simons terms
\be
H = d B - \cA \wedge F,
\ee
where $F = d A$.

Omitting a total divergence one obtains the following
four--dimensional action
\bea
S_4 &=& \int d^4x \sqrt{-g_4} \Big\{ R_4
  - \frac{16+3\alpha^2}{18\alpha^2}(\partial \phi)^2 \nonumber \\
 &-& \frac{8-3\alpha^2}{9\alpha^2}(\partial \phi)(\partial \psi)
  - \frac{4+3\alpha^2}{18\alpha^2}(\partial \psi)^2 \nonumber \\
 &-& \frac14 e^{\psi-\phi} \cF^2 - \frac14 e^{-\psi-\phi} F^2
  - \frac{e^{-2\phi}}{12} H^2 \Bigr\},
\eea
where
\be
\phi = \frac1{2} (\alpha \hat\phi + 4 \varphi), \qquad
\psi = \frac1{2} (\alpha \hat\phi -8 \varphi).
\end{equation}
The value of the dilaton coupling constant here is assumed to correspond
to dimensional reduction of the eleven--dimensional supergravity:
$\alpha^2=8/3$.
Dualising the three--form
\be
e^{-2\phi} H = - {}^* d \kappa,
\ee
one can write the corresponding action  as
\bea
S_4 &=& \int d^4x \sqrt{-g} \Big\{ R
  - \frac12 (\partial \phi)^2 - \frac14 (\partial \psi)^2 \nonumber \\
 &-& \frac12 e^{2\phi} (\partial \kappa)^2
  - \frac14 e^{\psi-\phi}\cF^2 - \frac14 e^{-\psi-\phi}F^2 \nonumber \\
 &-& \frac{\kappa}{4}\left( F{\tilde \cF} + \cF{\tilde F} \right) \Big\},
  \label{action}
\eea
where
\be
\tilde F_{\mu\nu} = \frac12 \sqrt{-g} \; \epsilon_{\mu\nu\lambda\rho}
       F^{\lambda\rho}, \quad
\tilde \cF_{\mu\nu} = \frac12 \sqrt{-g} \; \epsilon_{\mu\nu\lambda\rho}
       \cF^{\lambda\rho}.
\ee

Thus in four dimensions the theory is equivalent to Einstein gravity
coupled to two scalars, one pseudoscalar and two vector fields. The
equations of motion corresponding to this action consist of a coupled set
of Maxwell equations
\bea
\nabla_{\mu} && \left( e^{\psi-\phi}\cF^{\mu\nu}
  + \kappa \tilde F^{\mu\nu} \right) = 0, \\
\nabla_{\mu} && \left( e^{-\psi-\phi}F^{\mu\nu}
  + \kappa \tilde \cF^{\mu\nu} \right) = 0,
\eea
three equations for scalar fields
\bea
\nabla^2 && \phi - e^{2\phi} (\partial\kappa)^2
  + \frac14 e^{-\phi} \left( e^{\psi}\cF^2 + e^{-\psi}F^2 \right) = 0, \\
\nabla^2 && \kappa + 2 \partial_\mu \phi \partial^\mu \kappa
  - \frac14 e^{-2\phi} \left( \cF\tilde F + F\tilde\cF \right) = 0, \\
\nabla^2 && \psi - \frac12 e^{-\phi} \left(e^{\psi}\cF^2
  - e^{-\psi} F^2 \right) = 0,
\eea
and Einstein equations
\bea
R_{\mu\nu} && - \frac12 \partial_\mu \phi \partial_\nu \phi
  - \frac14 \partial_\mu \psi \partial_\nu \psi
  - \frac12 e^{2\phi} \partial_\mu \kappa \partial_\nu \kappa
  \nonumber \\
  && + \frac12 e^{\psi-\phi} \left(\cF_{\mu\alpha} \cF^\alpha{}_\nu
    + \frac14 g_{\mu\nu} \cF^2 \right)
  \nonumber \\
  && + \frac12 e^{-\psi-\phi} \left(F_{\mu\alpha} F^\alpha{}_\nu
    + \frac14 g_{\mu\nu} F^2 \right) = 0.
\eea
The Bianchi identities for two--forms read
\bea
\epsilon^{\mu\nu\lambda\rho} \partial_{\mu} F_{\nu\lambda} = 0, \quad
\epsilon^{\mu\nu\lambda\rho} \partial_{\mu} \cF_{\nu\lambda} = 0,
\eea

It is convenient to introduce the complex ``axidilaton''
\be
z = \kappa + i e^{-\phi},
\ee
and the following combination of the form fields:
\be
F_+ = e^{\psi/2} \cF + i e^{-\psi/2} \tilde F, \quad
F_- = e^{-\psi/2} F + i e^{\psi/2} \tilde \cF,
\ee
such that $F_+^2 = F_-^2$.
The action (\ref{action}) can be rewritten in the form
\bea
S_4 &=& \int d^x \sqrt{-g} \Big\{ R
  - \frac{|\partial z|^2}{2(\Im z)^2} - \frac14 (\partial \psi)^2
  \nonumber \\
 &-& \frac18 \Im \left[z(F_+^2 + F_-^2) \right] \Big\},
\eea
exhibiting the $SO(1,1)$ symmetry ($T$--duality)
\be
\psi \to \psi + \beta, \quad F_{\pm} \to F_{\pm}.
\ee
The field equations in terms of the complex variables
\bea
&&\nabla_{\mu} \Im\left( e^{\pm\psi/2}\,z\,F_{\pm}^{\mu\nu} \right)=0,\\
&&\nabla^2 z + i \frac{(\partial z)^2}{\Im z}
  - \frac{i}8 (\Im z)^2 ( \bar F_+^2 + \bar F_-^2 )=0, \\
&&\nabla^2 \psi - \frac14 \Im z ( |F_+|^2 - |F_-|^2 )=0, \\
&&R_{\mu\nu} - \frac{\partial_{(\mu} z \partial_{\nu)} \bar z}{2(\Im z)^2}
  - \frac14 \partial_\mu \psi \partial_\nu \psi \nonumber \\
&&\quad + \frac14 \Im z \Re \left(F_{+\mu\alpha} \bar F_+^\alpha{}_\nu
  + F_{-\mu\alpha} \bar F_-^\alpha{}_\nu \right) = 0,
\eea
together with Bianchi identities
\be
\nabla_{\mu} \Im \left( e^{\pm\psi/2}F_{\pm}^{\mu\nu} \right) = 0,
\ee
exhibit the $SL(2,R)$ $S$--duality symmetry
\be
z \to \frac{az+b}{cz+d}, \quad F_{\pm} \to (cz+d) F_{\pm},
\ee
where $ad-bc=1$.
Therefore, the duality group in four dimensions is
$SO(1,1)\times SL(2,R)$.

\section{Reduction to Three Dimensions}
As a second step we reduce the theory to three dimensions
assuming stationarity of the four-metric:
\be
ds_4^2 = -f (dt - \varpi_i dx^i)^2 + f^{-1} h_{ij} dx^i dx^j,
\ee
where $f, \varpi_i$ and $h_{ij}$ depend only on $x^i$.
In three dimensions vector fields can be parameterized by scalar potentials
$v_a, u_a, a=1,2$ via
\bea
&& \cF_{i0} = \partial_i v_1, \quad
  e^{\psi-\phi} \cF^{ij} + \kappa \tilde F^{ij}
    = \frac{-f}{\sqrt{h}} \epsilon^{ijk} \partial_k u_1, \\
&& F_{i0} = \partial_i v_2, \quad
  e^{-\psi-\phi} F^{ij} + \kappa {\tilde \cF}^{ij}
    = \frac{-f}{\sqrt{h}} \epsilon^{ijk} \partial_k u_2.
\eea
The three--dimensional KK field $\varpi_i$ then is dualized as follows
\be
\tau^i = \frac{f^2}{\sqrt{h}} \epsilon^{ijk} \partial_j \varpi_k.
\ee
The action for new variables reads
\bea
S_3 &=& \int d^3x \sqrt{h} \Big\{ R_3
  - \Big[ \frac1{2f^2} \left( (\partial f)^2 + \tau^2 \right)
  + \frac12 (\partial \phi)^2 \nonumber \\
 &+& \frac14 (\partial \psi)^2 + \frac12 e^{2\phi} (\partial \kappa)^2
  - \frac1{2f} \Big( e^{\psi-\phi} (\partial v_1)^2
    \!+\! e^{-\psi+\phi} w_1^2 \nonumber \\
 &+& e^{-\psi-\phi} (\partial v_2)^2
  + e^{\psi+\phi} w_2^2 \Big) \Big] \Big\},
\eea
where
\be
w_1 = \partial u_1 - \kappa \partial v_2, \quad
w_2 = \partial u_2 - \kappa \partial v_1,
\ee
and all vector contractions correspond to the metric $h_{ij}$.

One can solve the four--dimensional constraint equations for the Ricci
component $R^i_0$ by introducing the twist potential $\chi$
\be
\tau_i = \partial_i \chi + \frac1{2} \left( v_a \partial_i u_a
       - u_a \partial_i v_a \right),
\ee
so finally we obtain a three--dimensional $\sigma$--model with the
nine--dimensional target space parameterized by
$\Phi^A = (f, \chi, \phi, \psi, \kappa, v_1, u_1, v_2, u_2),
 A = 1, \ldots, 9$ and endowed with the metric
\bea
dl^2 &=& \frac1{2f^2} \left\{ df^2 + \left[ d\chi
  + \frac12 \left( v_a du_a - u_a dv_a \right) \right]^2 \right\}
    \nonumber \\
 &+& \frac12 d\phi^2 + \frac14 d\psi^2
  + \frac12 e^{2\phi} d\kappa^2 \nonumber \\
 &-& \frac1{2f} \Big[ e^{\psi-\phi} dv_1^2
  + e^{-\psi+\phi} (du_1 - \kappa dv_2)^2 \nonumber \\
 &+& e^{-\psi-\phi} (dv_2)^2 + e^{\psi+\phi} (du_2 - \kappa dv_1)^2
    \Big].  \label{TSM}
\eea
This is the metric of the symmetric space $SL(4,R)/SO(2,2)$ on which
the $SL(4,R)$ isometry group acts transitively. As the coset representatives
one can choose the symmetric $SL(4,R)$ matrices and finally the target space
metric will read
\be \label{TSM'}
dl^2 = -\frac14 \Tr \left( d {\cal M} d {\cal M}^{-1} \right).
\ee
The matrix ${\cal M}$ can be chosen in the form
\bea
{\cal M} &=&
  \left( \begin{array}{cc} 1 & 0 \\ Q^T & 1 \end{array} \right)
  \left( \begin{array}{cc} P_1^{-1} & 0 \\ 0 & P_2 \end{array} \right)
  \left( \begin{array}{cc} 1 & Q \\ 0 & 1 \end{array} \right),
  \nonumber \\
 &=& \left( \begin{array}{cc}
   P_1^{-1} & P_1^{-1} Q \\ Q^T P_1^{-1} & P_2 + Q^T P_1^{-1} Q
   \end{array} \right),
\eea
where the $P_1, P_2 $ and $Q$ are the real $2 \times 2$ matrices
and $P_1, P_2$ are symmetric matrices with the same determinant.
This matrix is a direct generalization of the $Sp(4,R)$ matrix found
in \cite{GaKe96} in the case of the four--dimensional EMDA theory.
Its inverse is given by
\be
{\cal M}^{-1} =
  \left( \begin{array}{cc}
   P_1 + Q P_2^{-1} Q^T & -Q P_2^{-1} \\ -P_2^{-1} Q^T & P_2^{-1}
  \end{array} \right).
\ee
The following identity is useful in establishing the equivalence
between (\ref{TSM}) and (\ref{TSM'}):
\bea
&& \Tr \left( d {\cal M} d {\cal M}^{-1} \right) \nonumber\\
 \equiv && \Tr \left( dP_1 dP_1^{-1} + dP_2 dP_2^{-1}
  - 2 dQ P_2^{-1} dQ^T P_1^{-1} \right).
\eea
Using it one can find the following parameterization of
the desired $2\times 2$ matrices in terms of the potentials:
\bea
P_1 &=& e^{\psi/2} \left( \begin{array}{cc}
      f e^{-\psi} - (v_1)^2 e^{-\phi} & -v_1 e^{-\phi} \\
      -v_1 e^{-\phi} & -e^{-\phi}
      \end{array} \right), \\
P_2 &=& e^{-\psi/2} \left( \begin{array}{cc}
      f e^\psi - (v_2)^2 e^{-\phi} & -v_2 e^{-\phi} \\
      -v_2 e^{-\phi} & -e^{-\phi}
      \end{array} \right),\\
Q &=& \left( \begin{array}{cc}
      \frac1{2} \xi-\chi & u_2 - \kappa v_1 \\
      u_1 - \kappa v_2 & -\kappa
      \end{array} \right),
\eea
where
\be
\xi = v_1 \left( u_1 - \kappa v_2 \right)
    + v_2 \left( u_2 - \kappa v_1 \right).
\ee
Therefore, the three--dimensional action can be rewritten as
\be
S_3 = \int d^3x \sqrt{h} \left\{ R_3
    + \frac14 \Tr \left( \nabla{\cal M} \nabla{\cal M}^{-1} \right)
      \right\}.
\end{equation}

This action is invariant under the 15--parametric $SL(4,R)$
transformations ${\cal M} \to {\cal G}^T {\cal M} {\cal G}$ with
constant ${\cal G}\in SL(4,R)$.
For ${\cal G}$ it is convenient to use the Gauss decomposition into
the product of {\it left} (triangle), {\it center} (block--diagonal)
and {\it right} (triangle) matrices as follow
\bea
&&{\cal G} = {\cal L} {\cal S} {\cal R}, \\
{\cal L} = \left( \begin{array}{cc} 1 & 0 \\ L & 1
                   \end{array} \right), \;
&&{\cal S} = \left( \begin{array}{cc} S^{-1} & 0 \\ 0 & T
                   \end{array} \right), \;
{\cal R} = \left( \begin{array}{cc} 1 & R \\ 0 & 1
                   \end{array} \right),
\eea
where $L, R$ are arbitrary real $2\times 2$ matrices, $S, T$ are
arbitrary real $2\times 2$ matrices satisfying $ \det S = \det T $.

It is easy to check that  $R$-transformations
\be
P \to P, \qquad F \to F, \qquad Q \to Q + R. \label{RT}
\ee
are pure gauge.
$S$-transformations
\begin{equation}
P_1 \to S P_1 S^T, \qquad    P_2 \to T^T P_2 T, \qquad    Q \to S Q T.
\end{equation}
contain two gauge, three different scale and two (electric type)
Harrison transformations. This subgroup is seven-parametric
and the corresponding parameter matrices read
\be
S = \left( \begin{array}{cc} e^{s_1} & s_2 \\ s_3 & e^{s_4}
     \end{array} \right), \quad
T = \left( \begin{array}{cc} e^{t_1} & t_2 \\ t_3 & e^{t_4}
     \end{array} \right)
\ee
where $e^{s_1+s_4}-s_2 s_3=e^{t_1+t_4}-t_2 t_3$.
One can identify the following one--parametric transformations

{\it Scale transformation} ($t_1=s_1$)
\be
  f \to e^{2s_1}f, \; \chi \to e^{2s_1} \chi, \;
  v_a \to e^{s_1} v_a, u_a \to e^{s_1} u_a, \label{t1}
\ee

{\it Dilaton shift} ($t_4=s_4$)
\be
\phi \to \phi - 2s_4, \; \kappa \to e^{2s_4} \kappa, \;
  v_a \to e^{-s_4} v_a, \; u_a \to e^{s_4} u_a, \label{t4}
\ee

{\it Modulus shift} ($-s_1=s_4=t_1=-t_4=t/2$)
\bea
&& \psi \to \psi + 2t, \nonumber \\
&& (v_1, u_2) \to e^{-t} (v_1, u_2), \quad
   (u_1, v_2) \to e^{t} (u_1,v_2), \label{t}
\eea

{\it Electric gauge}
\bea
  v_1 &\to& v_1 + s_2, \quad \chi \to \chi - \frac{s_2}{2} u_1, \label{s2} \\
  v_2 &\to& v_2 + t_3, \quad \chi \to \chi - \frac{t_3}{2} u_2. \label{t3}
\eea

There are two {\it electric Harrison} transformations: one generating
the Kaluza--Klein vector
\bea
&& \phi \to \phi - \frac12 \ln H_1, \quad \psi \to \psi + \ln H_1, \quad
   f \to f H_1^{-1/2}, \nonumber \\
&& v_1 \to \frac{H_2}{H_1}, \quad
   u_1 \to u_1 - s_3 \chi + \frac{s_3}{2}(v_1 u_1 - v_2 u_2), \nonumber \\
&& u_2 \to \frac1{H_1} \left[ u_2(1+s_3 v_1)
           -s_3\kappa f e^{-\psi+\phi} \right], \nonumber \\
&& \kappa \to \kappa - s_3(u_2 - \kappa v_1), \nonumber \\
&& \chi \to \chi\!+\!\frac{s_3 H_2}{4 H_1}(v_2 u_2\!-\!v_1 u_1\!-\!2\chi)
   \!-\! \frac{s_3 f e^{-\psi+\phi}}{2 H_1}(u_1\!-\!\kappa v_2), \nonumber \\
&& \hbox{where} \nonumber \\
&& \qquad H_1 = (1+s_3 v_1)^2 - s_3^2 f e^{-\psi+\phi}, \nonumber \\
&& \qquad H_2 = v_1(1+s_3 v_1) - s_3 f e^{-\psi+\phi}, \label{s3}
\eea
and another generating the three--form electric component
\bea
&& \phi \to \phi - \frac12 \ln H'_1, \quad \psi \to \psi - \ln H'_1, \quad
   f \to f H'_1{}^{-1/2}, \nonumber \\
&& v_2 \to \frac{H'_2}{H'_1}, \quad
   u_2 \to u_2 - t_2 \chi + \frac{t_2}{2}(v_2 u_2 - v_1 u_1), \nonumber \\
&& u_1 \to \frac1{H'_1} \left[ u_1(1+t_2 v_2)
           -t_2\kappa f e^{\psi+\phi} \right], \nonumber \\
&& \kappa \to \kappa - t_2(u_1 - \kappa v_2), \nonumber \\
&& \chi \to \chi\!+\!\frac{t_2 H'_2}{4H'_1}(v_1 u_1\!-\!v_2 u_2\!-\!2\chi)
   \!-\! \frac{t_2 f e^{\psi+\phi}}{2 H'_1} (u_2-\kappa v_1), \nonumber \\
&& \hbox{where} \nonumber \\
&& \qquad H'_1 = (1+t_2 v_2)^2 - t_2^2 f e^{\psi+\phi}, \nonumber \\
&& \qquad H'_2 = v_2(1+t_2 v_2) - t_2 f e^{\psi+\phi}. \label{t2}
\eea

The subgroup of the left transformations ${\cal L}$
is more difficult to write in terms of its action on the potentials.
In terms of $2\times 2$ matrices it reads:
\bea
&& P_1^{-1} \to (1 + QL)^T P_1^{-1} (1 + QL) + L^T P_2 L, \nonumber \\
&& P_2 + Q^T P_1^{-1} Q \to P_2 + Q^T P_1^{-1} Q, \nonumber \\
&& P_1^{-1} Q \to (1+L^T Q^T) P_1^{-1} Q + L^T P_2. \label{LT}
\eea
Using the parameterization
$L=\left( \begin{array}{cc} l_1 & l_2 \\ l_3 & l_4 \end{array} \right)$,
one can identify $l_1$ as a parameter of the Ehlers transformation,
$l_2,\, l_3$ as parameters of the magnetic Harrison transformations,
and $l_4$ as a parameter of the Ehlers--like part of $S$--duality.
The first three transformations are
rather complicated, so we give here explicitly only the last one:
\bea
&&z^{-1}\to z^{-1}+ l_4 \nonumber\\
&&v_1 \to v_1 - l_4 u_2, \qquad v_2 \to v_2 - l_4 u_1.
\eea
However, if the seed solution satisfies the four--dimensional vacuum
Einstein equations ({\em i.e.} corresponds to the $SL(2,R)/SO(2)$
subspace of the full target space), the $2\times 2$ matrices admit
a simple form
\be
P_1 = P_2 = \left( \begin{array}{cc}
    f_0 & 0 \\ 0 & -1 \end{array} \right), \qquad
Q = \left( \begin{array}{cc}
    -\chi_0 & 0 \\ 0 & 0 \end{array} \right),
\ee
and the left transformations considerably simplify.
The Ehlers transformation will take the form
\be
f = \frac{f_0}{(1-l_1\chi_0)^2 + l_1^2 f_0^2}, \;
\chi=\frac{\chi_0(1-l_1\chi_0)-l_1 f_0^2}{(1-l_1\chi_0)^2+l_1^2 f_0^2},
\ee
the magnetic Harrison transformation for the KK vector will be
\bea
\phi &=& \frac12 \ln(1-l_2^2 f_0), \qquad
  \psi = -\ln(1-l_2^2 f_0), \nonumber \\
f &=& f_0 (1-l_2^2 f_0)^{-\frac12}, \qquad
  \chi = \chi_0 \left[ 1 + \frac{l_2^2 f_0}{2(1-l_2^2 f_0)} \right],
  \nonumber \\
u_1 &=& \frac{-l_2 f_0}{1-l_2^2 f_0}, \qquad v_1 = l_2 \chi_0,
\eea
while the magnetic Harrison transformation generating the three--form field
can be expressed as
\bea
\phi &=& \frac12 \ln(1-l_3^2 f_0), \qquad
  \psi = \ln(1-l_3^2 f_0), \nonumber \\
f &=& f_0 (1-l_3^2 f_0)^{-\frac12}, \qquad
  \chi = \chi_0 \left[ 1 + \frac{l_3^2 f_0}{2(1-l_3^2 f_0)} \right],
  \nonumber \\
u_2 &=& \frac{-l_3 f_0}{1-l_3^2 f_0}, \qquad v_2 = l_3 \chi_0.
\eea

Two other useful subspaces of the full target space (\ref{TSM})
correspond to the five--dimensional vacuum ($v_2=u_2=\kappa=0, \psi=-2\phi$,
the symmetry group being $SL(3,R)$)
and to the four-dimensional EMDA system ($v_2=v_1, u_2=u_1,
\psi=0$, with the symmetry $Sp(4,R)$). The corresponding solutions can
be used as seeds as well.

\section{Solution Generation}

Our interpretation of the $SL(4,R)$ transformations is based on the
four-dimensional picture, therefore two type Harrison transformations
correspond to different five-dimensional charges: one pair is related
to the non-diagonal components of the five-dimensional metric,
while another pair --- to the three-form field. Thus, taking as seeds the
four-dimensional vacuum   solutions, one can generate in the first case
the vacuum 5D metrics, and in the second --- the three-form
supported configurations. Starting with the $y$--translated
four-dimensional Schwarzschild solution
\bea
ds_5^2 &=& -(1-\frac{2m}{r}) dt^2 + dy^2 \nonumber \\
       &+& (1-\frac{2m}{r})^{-1} dr^2 + r^2 d\Omega_2,
\eea
and applying the  sequences of transformations including the second type
electric Harrison transformation one arrives
at the {\it electric string solution}~\cite{DuKhLu95}. More
precisely, the three--form electric component is produced by the
electric Harrison transformations (\ref{t2}) with the parameter $t_2=\mu$,
then one has to compensate the asymptotic values of variables
to make the solution asymptotically flat.
One possibility is to use the following  chain of transformations:
the electric gauge (\ref{t3}), $t_3=\mu/(1-\mu^2)$;
the scale transformation (\ref{t1}), $t_1=s_1=\frac14\ln(1-\mu^2)$;
the dilaton shift (\ref{t4}), $t_4=s_4=-\frac14\ln(1-\mu^2)$ and
the modulus shift (\ref{t}), $t=\frac12\ln(1-\mu^2)$.
The final result is
\bea  \label{elst}
ds_5^2 &=& (1+\frac{2q}{r})^{-1/3} \left\{ -(1-\frac{2m}{r}) dt^2
        + dy^2 \right\} \nonumber \\
       &+& (1+\frac{2q}{r})^{2/3} \left\{ (1-\frac{2m}{r})^{-1} dr^2
        + r^2 d\Omega_2 \right\}, \nonumber \\
\hat\phi &=& -\sqrt{\frac23} \ln (1+\frac{2q}{r}), \nonumber \\
\hat B_{ty} &=& \sqrt{\frac{q}{m+q}} \left[ 1 - (1-\frac{2m}r)
        (1+\frac{2q}r)^{-1} \right], \label{ss1}
\eea
where
\be
q := \frac{\mu^2 m}{1-\mu^2}.
\ee
This is the black string, of which the extremal version is achieved
in the limit $\mu\to 1, m\to 0, q$ finite.

One can apply a similar chain of transformations including the second
type magnetic Harrison transformation (\ref{LT}) with  $l_3=\mu$. The
necessary adjustment includes
the magnetic gauge (\ref{RT}) ($r_2=\mu/(1-\mu^2)$),
the scale transformation (\ref{t1}) ($t_1=s_1=\frac14\ln(1-\mu^2)$),
the dilaton shift (\ref{t4}) ($t_4=s_4=\frac14\ln(1-\mu^2)$) and
the modulus shift (\ref{t}) ($t=-\frac12\ln(1-\mu^2)$). The resulting
solution has the form
\bea \label{mst}
ds_5^2 &=& -(1+\frac{2p}r)^{-2/3} (1-\frac{2m}r) dt^2 \nonumber \\
       &+& (1+\frac{2p}{r})^{1/3} \left\{ dy^2+(1-\frac{2m}{r})^{-1}
           dr^2 + r^2 d\Omega_2 \right\}, \nonumber \\
\hat\phi &=& \sqrt{\frac23} \ln (1+\frac{2p}r), \nonumber \\
\hat B_{\phi y} &=& -2 \sqrt{p(m+p)} \cos\theta, \label{ss0}
\eea
where
\be
p := \frac{\mu^2 m}{1-\mu^2}.
\ee
To interpret it physically, one has to note that in five dimensions
the three-form is dual to the two-form field $^*H$, which will have non-zero
the components $^*H_{tr}$ like in the case of the four-dimensional
magnetically charged black hole. Thus we deal with the black 5D
{\em magnetic 0-brane}, of which the extremal version corresponds to
$m=0$.

Application of the first type Harrison transformations generate
vacuum 5D solutions. Note, that a variety of such solutions were
found in the past using different
techniques~\cite{DoMa82,GrPe83,So83,Cl86,GiWi86,FrZeBl87,GiMa88,Ra95}.
All of them, as well as their generalizations including  a NUT
parameter, can be found via suitable chains of the $SL(3,R)$ transformations
(forming a subgroup of the $SL(4,R)$ with $v_2=u_2=\kappa=0,\psi=-2\phi$)
from the four-dimensional Kerr metric. For more detailed analysis
of the vacuum 5D metrics with two commuting isometries through the
the $SL(3,R)$ symmetry see~\cite{Cl86}.

One is got by an
electric Harrison transformation (\ref{s3}) with a parameter $s_3=\mu$
together with an appropriate compensation of asymptotic values:
\bea
ds_5^2 &=& -(1+\frac{2q}r)^{-1} (1-\frac{2m}{r}) dt^2 \nonumber\\
  &+& (1+\frac{2q}r) \left[ dy + 2\sqrt{q(m+q)} (r+2q)^{-1} dt \right]^2
  \nonumber \\
  &+& (1-\frac{2m}r)^{-1} dr^2 + r^2 d\Omega_2.
\eea
This is the black version of the Dobiasch--Maison
solution~\cite{DoMa82}, the original form of which is recovered when $m=0$ .

Another 5D vacuum solution resulting from the first type magnetic
Harrison transformation (\ref{LT}) with
a parameter $l_2=\mu$ is
\bea
ds_5^2 &=& -(1-\frac{2m}r) dt^2 \nonumber \\
  &+& (1+\frac{2p}r)^{-1} \left[ dy
   + 2 \sqrt{p(m+p)} \cos\theta d\phi \right]^2 \nonumber \\
  &+& (1+\frac{2p}r) \left[ (1-\frac{2m}r)^{-1} dr^2
   + r^2 d\Omega_2 \right].
\eea
This is the black version of the Gross--Perry--Sorkin monopole
\cite{GrPe83,So83}, the original form corresponding to $m=0$.

The total number of the independent conserved charges in the present theory
is seven: mass, NUT, four charges associated with $F$ and $\cal F$, and one
rotation parameter. Note that, since
we are considering the class of 5D metrics with non-compact Killing orbits,
only one rotation parameter is allowed. Therefore, the general string
solution should contain seven free parameters. Here we exhibit only the
simplest four-parametric subfamilies which can be presented in a concise form.

One can repeat the same calculations as before using as a seed the
$y$-translated four--dimensional vacuum Kerr--NUT metric
\bea
ds_5^2 &=& -\frac{\Delta-a^2\sin^2\theta}{\Sigma}(dt-\omega d\phi)^2
  + dy^2 \nonumber  \\
 &+& \Sigma \left( \frac{d r^2}{\Delta} + d\theta^2
  + \frac{\Delta\sin^2\theta}{\Delta-a^2\sin^2\theta}d\phi^2 \right),
\eea
where
\bea
\Delta &:=& r^2 - 2mr + a^2 - n^2, \\
\Sigma &:=& r^2 + (a\cos\theta+n)^2, \\
\omega &:=& \frac{2n\Delta\cos\theta + 2a\sin^2\theta(mr+n^2)}
                 {a^2\sin^2\theta-\Delta}.
\eea
This family contains three free parameters  $m,a,n$ (mass, rotation and NUT)
and the corresponding non-zero seed potentials read
\bea
f_0 &=& \frac{r^2-2mr+a^2\cos^2\theta-n^2}{r^2+(a\cos\theta+n)^2},
  \nonumber \\
\chi_0 &=& \frac{2m(a\cos\theta+n)-2nr}{r^2+(a\cos\theta+n)^2}.
\eea

Applying the second type electric Harrison chain one obtains
\bea
ds_5^2 &=& H^{-\frac13}\left[dy^2 - \frac{\Delta-a^2\sin^2\theta}{\Sigma}
  (dt - \cosh\delta\;\omega d\phi)^2 \right] \nonumber \\
  &+& H^\frac23 \Sigma \left( \frac{dr^2}{\Delta} + d\theta^2
   + \frac{\Delta\sin^2\theta}{\Delta-a^2\sin^2\theta}d\phi^2 \right),
   \nonumber \\
\hat \phi &=& -\sqrt{\frac23} \ln H, \nonumber \\
\hat B_{ty} &=& \sinh 2\delta (mr+an\cos\theta+n^2)(\Sigma H)^{-1},
    \nonumber \\
\hat B_{\phi y} &=& \!-\! 2\sinh\!\delta [n\Delta\cos\theta
  \!+\! a\sin^2\theta(mr\!+\!n^2)](\Sigma H)^{-1}\!,
\eea
where $\delta$ is a three-form charge parameter, and
\be
H := 1 + 2\sinh^2\delta (mr+an\cos\theta+n^2)\Sigma^{-1}.
\ee
This is a NUT generalization of the black rotating string found
previously in~\cite{HoHo92}.

Similarly, applying a magnetic second type Harrison transformation
one obtains a four-parametric family of rotating magnetic zero-branes:
\bea
ds_5^2 &=& -H^{-\frac23} \frac{\Delta-a^2\sin^2\theta}{\Sigma}
  (dt - \cosh\delta\;\omega d\phi)^2 \nonumber \\
  &+& H^\frac13 \left[dy^2 + \Sigma\left(\frac{dr^2}{\Delta}+d\theta^2
   + \frac{\Delta\sin^2\theta}{\Delta-a^2\sin^2\theta}d\phi^2 \right)
     \right], \nonumber \\
\hat \phi &=& \sqrt{\frac23} \ln H, \nonumber \\
\hat B_{ty} &=& 2\sinh\delta [m(a\cos\theta+n)-nr]\Sigma^{-1},
     \nonumber \\
\hat B_{\phi y} &=& -\sinh 2\delta \Big\{m\cos\theta \nonumber\\
  &+&\!(2n\cos\theta\!-\!a\sin^2\theta)[nr\!-\!m(a\cos\theta\!+\!n)]
     \Sigma^{-1}  \Big\}.
\eea

Application of the first type Harrison transformations gives
the NUT generalization of the rotating solution found in ~\cite{FrZeBl87}
(which is the rotating version of the Dobiasch-Maison solution):
\bea
ds_5^2 &=& H \left(dy + \cA_t dt + \cA_\phi d\phi\right)^2 \nonumber \\
  &-& H^{-1} \frac{\Delta-a^2\sin^2\theta}{\Sigma}
      (dt - \cosh\delta\;\omega d\phi)^2  \nonumber \\
  &+& \Sigma \left( \frac{d r^2}{\Delta} + d\theta^2
   + \frac{\Delta\sin^2\theta}{\Delta-a^2\sin^2\theta}d\phi^2 \right),
     \nonumber \\
\cA_t &=& \sinh 2\delta (mr+an\cos\theta+n^2)(\Sigma H)^{-1},
    \nonumber \\
\cA_\phi &=& \!-\!2\sinh\delta [n\Delta\cos\theta
  \!+\! a\sin^2\theta(mr\!+\!n^2)](\Sigma H)^{-1},
\eea
and the rotating generalization (with NUT) of the Gross-Perry-Sorkin
monopole:
\bea
ds_5^2 &=& H^{-1} \left(dy + \cA_t dt + \cA_\phi d\phi\right)^2 \nonumber\\
  &-& \frac{\Delta-a^2\sin^2\theta}{\Sigma}
      (dt - \cosh\delta\;\omega d\phi)^2 \nonumber \\
  &+& H \Sigma \left( \frac{d r^2}{\Delta} + d\theta^2
   + \frac{\Delta\sin^2\theta}{\Delta-a^2\sin^2\theta}d\phi^2 \right),
     \nonumber \\
\cA_t &=& 2\sinh\delta [m(a\cos\theta+n)-nr]\Sigma^{-1},
     \nonumber \\
\cA_\phi &=& -\sinh 2\delta \Big\{m\cos\theta \nonumber\\
  &+&\!(2n\cos\theta\!-\!a\sin^2\theta)[nr\!-\!m(a\cos\theta+n)]
     \Sigma^{-1} \Big\}.
\eea

\section{$(0-1)$ branes}
Combining together electric and magnetic Harrison transformations one
can obtain solutions with larger number of free parameters. From these
we give explicitly  a relatively simple composite
string --  magnetic zero-brane solution. It can be derived applying
a sequence of the second type electric and magnetic Harrison transformations
to the seed Schwarzschild metric. The resulting solution is
\bea
ds_5^2 &=& - \left( 1-\frac{2q}{r} \right)^{-\frac13}
             \left( 1+\frac{2p}{r} \right)^{-\frac23}
             ( dt\!+\!2\sqrt{pq}\cos\theta d\phi)^2  \nonumber \\
       &+& \left( 1-\frac{2q}{r} \right)^{-\frac13}
           \left( 1+\frac{2p}{r} \right)^{\frac13} dy^2 \nonumber \\
       &+& \left( 1-\frac{2q}{r} \right)^{\frac23}
           \left( 1+\frac{2p}{r} \right)^{\frac13}
           ( dr^2 + r^2 d\Omega_2 ), \nonumber \\
\hat \phi &=& \sqrt{\frac23} \left[ \ln \left( 1+\frac{2p}r \right)
              - \ln \left( 1-\frac{2q}r \right) \right],
           \nonumber \\
B_{ty} &=& \sqrt{\frac{q}{p+q}} \left[ 1 - \left( 1+\frac{2p}r \right)
           \left( 1-\frac{2q}r \right)^{-1} \right], \nonumber \\
B_{\phi y} &=& -2\sqrt{p(p+q)} \left( 1-\frac{2q}r \right)^{-1}
           \cos\theta.
\eea
When $p=0$ the solution reduces to the electric string (\ref{elst}),
while for  $q=0$ one recovers the magnetic zero-brane (\ref{elst}).
Note that, when both charges are non-zero, one has also a NUT parameter,
so the solution is not strictly asymptotically flat. One can compensate
the NUT charge via the Ehlers transformation thus making it an additional
free parameter. We intend to discuss this and more general solutions in
a separate publication.

\section{Concluding remarks}
We have shown that large classes of solutions to 5D dilaton-axion
gravity admitting two commuting isometries with non-compact orbits
can be obtained in a unique way using the $SL(4,R)$ duality from the
vacuum four-dimensional metrics such as Schwarzschild or Kerr-NUT.
This class does not include spherically symmetric or rotating
five-dimensional black holes which, although possess two commuting Killing
symmetries too, have one of them of the rotational type.
These solutions are still
within the scope of the present $\sigma$-model, but the group
transformations generically relate them to asymptotically non-flat metrics
(e.g. corresponding to black holes in the field of a fluxbrane
\cite{GaRy98}) and we did not consider them here.
The $SL(4,R)$ duality also connects solutions of the 5D dilaton-axion gravity
with solutions of the four-dimensional
EMDA system, which constitute a subspace ($v_1=v_2, u_1=u_2, \psi=0$)
of the full target space and may be used as seeds to generate five-dimensional
solutions.

An essentially new solution we presented here is the composite
$(0-1)$-brane which is a superposition of an electric string and
a $0$-brane supported by the magnetic sector of the three-form field.
This is an example of a dyonic composite solution in the case
when electric and magnetic branes have different world-volume dimensions.
The situation is similar to that in the eleven-dimensional supergravity
where one has electric two-branes and magnetic five-branes. In the latter case
composite electric-magnetic solutions are realized either as intersecting
two- and five-branes, or as composite $2\subset 5$ branes \cite{Co96}.
Note that the five-dimensional lagrangian does not contain
the Chern-Simons term which is essential for existence of the
$2\subset 5$ brane in the eleven-dimensional supergravity.
\section*{Acknowledgments}
One of the authors (DVG) would like to thank the Yukawa Institute for
Theoretical Physics, Kyoto, where the part of this work was done.
Fruitful discussions with T.~Nakamura and M.~Sasaki are gratefully
acknowledged. The research was supported in part by
the RFBR Grant 96--02--18899.


\end{multicols}
\end{document}